# Acoustic excitation of superharmonic capillary waves on a meniscus in a planar micro-geometry

Jie Xu and Daniel Attinger

Laboratory for Microscale Transport Phenomena, Department of Mechanical Engineering, Columbia University, New York, NY 10027

The effects of ultrasound on the dynamics of an air-water meniscus in a planar micro-geometry are investigated experimentally. The sonicated meniscus exhibits harmonic traveling waves or standing waves, the latter corresponding to a higher ultrasound level. Standing capillary waves with subharmonic and superharmonic frequencies are also observed, and are explained in the framework of parametric resonance theory, using the Mathieu equation.

Instabilities at a liquid-gas interface appear in a wide range of physical systems, from large ocean waves to microscopic oscillations of a cavitating bubble.[1,2] For a linear and conservative system the response is harmonic, i.e. with the same frequency as the excitation frequency.[3] However, in 1831 Faraday observed[4] the subharmonic response of a liquid-gas interface excited by vertical oscillations: the response frequency was half that of the excitation.[5] Faraday waves are caused by the excitation of the gravity term of their eigen frequency[5]. This phenomenon can be explained by a Mathieu equation[4], and has been widely used to generate waves in large systems,[2,4,6,7]. At smaller scale, gravity is negligible, and ultrasound is therefore more efficient for exciting waves. Using ultrasound, subharmonic responses have been observed[8-10] on millimeter-size bubbles, a phenomenon relevant for bubble sizing,[8] cavitation control,[9] and acoustic streaming.[10] Although the possibility of superharmonic response at a liquid-gas interface has been theoretically demonstrated[11-14], and superharmonic noise was recorded from a bubble cloud during a sonoluminescence study, no superharmonic wave at a liquid-gas interface has ever been directly observed. The occurrence of superharmonic waves was also qualified by Eisenmenger[13] as improbable because of the higher damping associated with higher frequencies. In our study, we excite with ultrasound a liquid-gas meniscus pinned at the junction between a microchannel and a chamber, in a microfluidic chip. We observe superharmonic oscillations at the meniscus, as well as harmonic and subharmonic oscillations. These phenomena are described and explained in the framework of parametric resonance theory.

A typical microfluidic chip used in our study is shown in Fig. 1(a): it involves a chamber (E) fed by a fork-like network of four channels A, B, C, D, with respective widths of 400, 1000, 400 and 100 μm. The height of each micro-channel is 50 μm. The microfluidic chip is manufactured in PDMS (polydimethylsiloxane) using the soft lithography process described in [15]. The PDMS chip is covered with a PDMS plate and sandwiched between two glass slides. This type of assembly ensures that every channel wall is made of PDMS, for the consistency of surface properties. A piezoelectric actuator is embedded in the PDMS cover plate on top of chamber $E$ to generate pressure waves in water. A stable air-water interface is then generated at the junction of channel $D$ and the chamber by filling the chamber with water and then injecting air in channel $D$. Note in Fig. 1(b) that two micro-geometrical features enhance the stability of the air-water interface[16]. The piezoelectric actuator is driven by a function generator (Agilent, 33120A), and a voltage amplifier (Krohn-Hite, 7600M), at frequencies up to 300kHz. Visualization is performed with a long-distance microscope in the plane perpendicular to the microfluidic chip, with a spatial resolution of 1 μm. A strobe microscopy technique is used for freezing the meniscus shape [17]. In this technique, a single function generator drives both the piezoelectric actuator and the strobe diode, while a delay generator controls the delay between the diode illumination and the actuation of the piezoelectric transducer. Finally, a frequency divider is used between the function generator and the photodiode, so that the diode frequency can be set to either the actuator frequency or half of its value, allowing the observation of superharmonic and subharmonic oscillations corresponding to half integer multiples of the excitation frequency. Deionized water is used in the experiments. The physical properties used in this study are described in Table I. The surface tension, density and viscosity are obtained from [18]. The contact angles are measured from the microscopy pictures.

| Symbol | Physical property | Value |
|--------|-------------------|-------|
| $\sigma$ | Surface tension of water | 72.0 mN/m @ 25 °C |
| | | 69.6 mN/m @ 40 °C |
| $\rho$ | Density of water | 0.998 g/cm$^3$ |
| $\theta$ | Contact angle of water on PDMS | 70°, receding |
| | | 110°, advancing |

TABLE I. Physical properties

In our experiments, the piezoelectric transducer induces pressure oscillations in the water with a frequency $f_e$, that excite interfacial waves at a frequency $f_1$. For relatively low excitation intensities and 100 kHz $< f_e <$ 170 kHz, we observed waves traveling on the meniscus surface at a frequency $f_1=f_e$. Fig. 2 shows a sequence taken at $f_e$=150 kHz, with a delay of 2 μs between each frame. Crests A and B are moving symmetrically towards the center of the channel, where they meet each other. This phenomenon occurs with the meniscus assuming either a curved shape (as in Fig. 2) or a flat shape (as in Fig. 5). Assuming conservative uncertainties $\Delta\lambda$ of $\pm 2$ μm for the wavelength, the observed wavelengths and oscillation frequencies are plotted in Fig. 3. Since the wavelengths are small ($\lambda \langle\langle 2\pi\kappa^{-1}$, where $\kappa^{-1}$ is the water-air capillary length, with $2\pi\kappa^{-1}$ typically on the order of a centimeter), gravity can be neglected in the theoretical analysis: very likely, these waves are capillary waves, caused by joint effect of inertia and surface tension.

In a 2D case, neglecting dissipative effects, the relationship between the wavelength and the oscillation frequency $f_0$ is given by the dispersion equation: [19]

$$\lambda_0 = \left(\frac{2\pi\sigma}{f_0^2 \rho}\right)^{\frac{1}{3}}. \qquad (1)$$

This relation is plotted in Fig. 3, for a water surface tension at the value of 40°C, corresponding to the measured water temperature, slightly heated by resistive dissipation in the piezoelectric transducer. The relatively good agreement in Fig. 3 between the experimental data and the 2D theory can be justified by the fact that the curvature in the observation plane (radius on the order of 10 µm) is much larger than the curvature in the perpendicular plane (radius on the order of 80 µm). It appears therefore that the proposed micro-channel/chamber configuration provides a simplified 2D model platform to observe wave at liquid-gas surfaces, without the complexity of configurations such as oscillating 3D bubbles in an unbounded fluid [9,20].

Increasing the actuation intensity –while keeping the excitation frequency $f_e$ constant at 150kHz- generates a standing wave, which is described by the successive frames in Fig. 4. The standing wave has a larger amplitude (around 5 µm) than the traveling wave (around 2 µm, see Fig. 2). Also, in the case of Figure 4, the frequency of the standing wave $f_1 = 1/2\, f_e$ is a subharmonic of the excitation frequency $f_e$. This phenomenon is also observed for a flat meniscus (Fig. 5). Typical experimental conditions with the threshold voltage above which standing waves replace traveling waves are given in Table II. Traveling waves are always harmonic at the excitation frequency $f_e$, while standing waves have a frequency $f_1$ that can be either subharmonic, harmonic, or superharmonic. For standing waves on a meniscus of length $L$ fixed on two lateral walls, the allowable wavelengths are

$$\lambda = \frac{L}{n+1} \qquad (2)$$

where *L* and *n* are respectively the channel width and a positive integer [21]. Fig. 5 shows that the system spontaneously selects *n*=1.

| $f_e$ (kHz) | $f_1/f_e$ | $f_1$ (kHz) | Threshold voltage needed to induce standing wave (V) |
|---|---|---|---|
| 175 | ½ | 87.5 | 98.6 |
| 150 | ½ | 75 | 29.9 |
| 127 | ½ | 63.5 | 77.4 |
| 75 | 1 | 75 | 182.8 |
| 74 | 1 | 74 | 182.3 |
| 50 | 3/2 | 75 | 107.7 |
| 49 | 3/2 | 73.5 | 95.1 |

**TABLE II. Conditions of the transition from traveling to standing waves**

The occurrence of subharmonic and superharmonic waves can be explained as the result of parametric oscillations [22]. Let's express a standing wave as the superposition of two waves traveling in opposite directions:

$$\eta = a\cos(\tfrac{2\pi}{\lambda}x - \omega t) + a\cos(\tfrac{2\pi}{\lambda}x + \omega t) = 2a\cos\tfrac{2\pi}{\lambda}x\cos\omega t \quad . \quad (3)$$

The resulting surface displacement therefore both obeys the wave equation and the oscillator equation, the latter being of the general form (neglecting damping)

$$\ddot{\eta} + \omega^2\eta = 0 \quad . \tag{4}$$

Fig. 6 describe the wave and reference frame in two planes, the top one parallel to the chip surface and the bottom one perpendicular to it, with the *x-z* plane parallel to the chip surface.

Assuming the motion is only in the *x-z* plane and the flow is irrotational, the velocity potential $\phi$ can be found by solving the Laplace equation

$$\nabla^2 \phi = 0, \tag{5}$$

with zero velocity boundary condition at infinity and a kinematic boundary condition on the meniscus:

$$u = \frac{\partial \phi}{\partial x} = 0 \quad w = \frac{\partial \phi}{\partial z} = 0 \quad \text{at} \quad z = \infty \tag{6}$$

$$\frac{\partial \eta}{\partial t} = \frac{\partial \phi}{\partial z} \quad \text{at} \quad z = 0 \tag{7}$$

In addition, a dynamic boundary condition arises from the Laplace pressure $\Delta P = \sigma \left( \frac{1}{R_h} + \frac{1}{R_w} \right)$ across the meniscus interface, where the curvature radii are calculated from the 3D meniscus shape shown in Fig. 6

$$\frac{1}{R_w} = \frac{-\partial^2 \eta / \partial x^2}{[1 + (\partial \eta / \partial x)^2]^{3/2}} \approx -\frac{\partial^2 \eta}{\partial x^2} \tag{8}$$

and

$$\frac{1}{R_h} = \frac{2(\eta - z_0)}{(\eta - z_0)^2 + h^2/4} \tag{9}$$

In the above equation, $z_0$ is defined in the legend of Fig. 6. For irrotational flow, the linearized Bernoulli equation is applicable,

$$\frac{\partial \phi}{\partial t} + \frac{p}{\rho} = 0 \tag{10}$$

Combining the above equation with the Laplace boundary condition gives a third boundary condition

$$\rho \frac{\partial \phi}{\partial t} = \sigma \left( \frac{1}{R_h} + \frac{1}{R_w} \right) \quad \text{at} \quad z = 0 \tag{11}$$

Solving Equation (5) with boundary conditions (6), (7) and (11) and assuming $z_0 = 0$ gives a dispersion equation between the wavelength $\lambda$ and frequency $f$

$$f^2 = \frac{\sigma}{\pi \rho \lambda^3} \left( 2\pi^2 + \frac{4\lambda^2}{h^2} \right). \tag{12}$$

Inserting the above equation into Equation (4), we obtain

$$\ddot{\eta} + \frac{16\pi\sigma}{\lambda \rho} \left( \frac{\pi^2}{2\lambda^2} + \frac{1}{h^2} \right) \eta = 0 \quad . \tag{13}$$

It is reasonable to assume that the chamber height $h$ is deforming with the piezoelectric deformation, with a small amplitude $\Delta h$ at the excitation frequency $f_e$. In that case, we can write $h = h_0 + \Delta h \cos(\omega_e t)$ in Equation (13). Neglecting second order terms, we obtain

$$\ddot{\eta} + \frac{16\pi\sigma}{\lambda \rho} \left( \frac{\pi^2}{2\lambda^2} + \frac{1}{h_0^2} - \frac{2\Delta h}{h_0^3} \cos(\omega_e t) \right) \eta = 0 \tag{14}$$

which is a Mathieu Equation [22] representing a parametrically excited oscillator. Although this equation cannot be solved analytically, the boundaries between stable and unstable solutions can be analytically determined using the formulas in [22] and are shown in Fig. 7. Fig. 7 shows for a case where damping is negligible the zones of stability and instability, the latter being shown by thin lines that correspond to the bottom of Mathieu tongues. States in the instability zones correspond to the possibility of standing waves, as shown in [11]. In the above derivation, the natural oscillation pulsation $\omega_0$ of our system was identified as the pulsation $\omega_1$ of the harmonic standing wave observed in Table II, which correspond to the $f_1$ = 75kHz case. The displacement $\Delta h$ of the microchannel height is measured as the piezoelectric transducer deformation using atomic force microscopy, to be on the order of 200nm. In Fig. 7, we have reported groups of experimental data where the transition from traveling to standing waves is observed during the process of slowly increasing the excitation frequency. The experimental points are gathered around the three instability zones, with groups of frequency ratios corresponding to the ones predicted by the Mathieu Equation. In Fig. 7, all the measured points close to the 1st order zone have $f_1/f_e$=1/2 (subharmonic response), while the points close the second and third order zone have $f_1/f_e$ equal respectively to 1 (harmonic) and 3/2 (superharmonic), as predicted theoretically[13,22]. In Fig. 7, the points corresponding to higher-order standing waves are higher on the vertical scale, which might be due to high order nonlinear damping. For instance, it has been shown by Insperger and Stepan[23], that adding viscous damping to the Mathieu treatment pulls the Mathieu tongues upwards. In any case, the general trend revealed by our experiments in Figure 7 is that instability states of order higher than ½ are more difficult to produce than the ½ order instability state. This might explain why subharmonic oscillations are easier to observe at a liquid-gas interface.

In summary, the response of a water-air meniscus to ultrasonic excitation has been studied. When the excitation is weak, traveling waves are found. When the excitation becomes stronger, standing waves appear at the meniscus interface. The ratio of the wave frequencies over the excitation frequencies assumes half integer values such as 1/2, 1, 3/2, a phenomenon explained by the Mathieu Equation. We thank the US National Science Foundation for their support through the CAREER grant 0449269. We thank Howard Stone (Harvard) and Michael Weinstein (Columbia APAM) for valuable suggestions.

**FIG. 1. (a) the microfluidic chip involves a chamber fed by a fork-like network of four channels; (b) detail of the meniscus. Triangular micro-features pin the meniscus wetting line.**

**FIG. 2. Traveling waves on meniscus at 150 kHz. The pictures at t>0 have artificially enhanced contrast.**

**FIG. 3. Wavelength vs wave frequency**

**FIG. 4.** Standing wave on meniscus interface excited at 150 kHz. The standing wave oscillates at 75 kHz. Note some condensation occurring along the walls of the air channel, a phenomenon found more often in the standing wave case than the traveling wave case.

**FIG. 5.** Standing waves at different frequencies.

**FIG. 6.** 3D shape of the meniscus oscillation, where $z_0$ is the distance between contact line and the $z$-location corresponding to $\eta = 0$.

**FIG. 7.** Instability analysis of Equation (14), where the instability zones are thin lines that correspond to the bottom of Mathieu tongues. The stars denote experimental results and the circled number near the data points equals $f_1/f_e$.

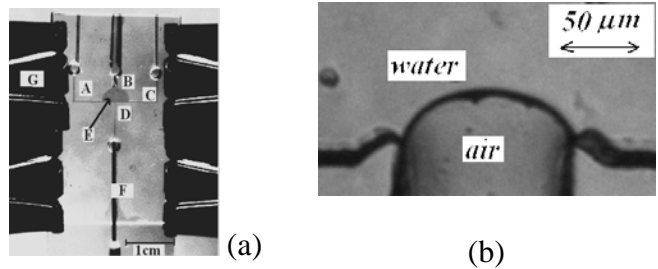

FIG. 1

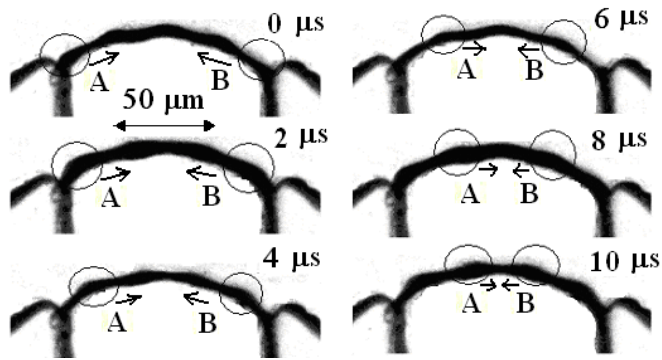

FIG. 2

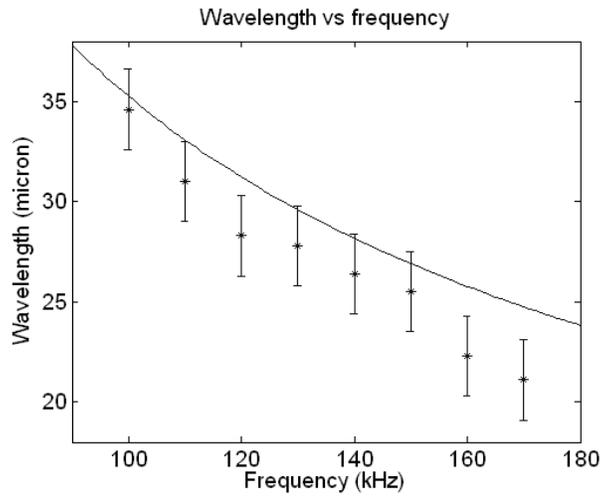

**FIG. 3**

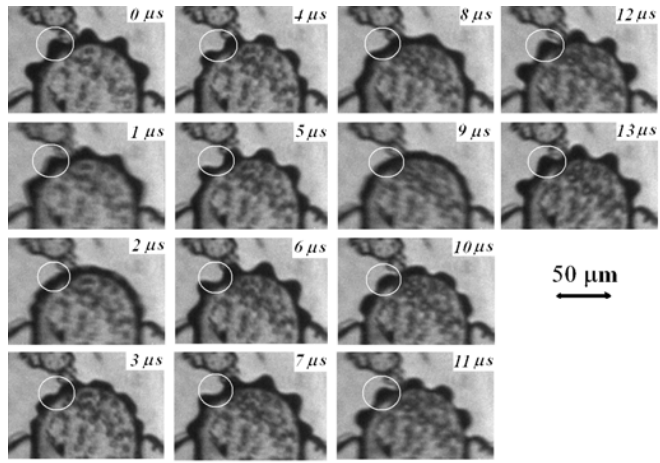

**FIG. 4**

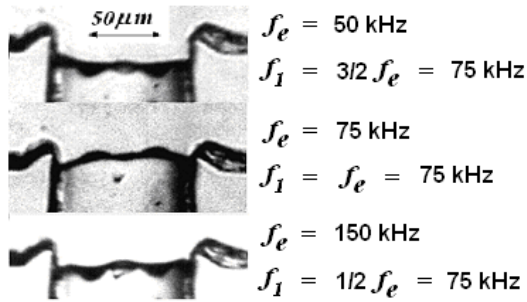

**FIG. 5**

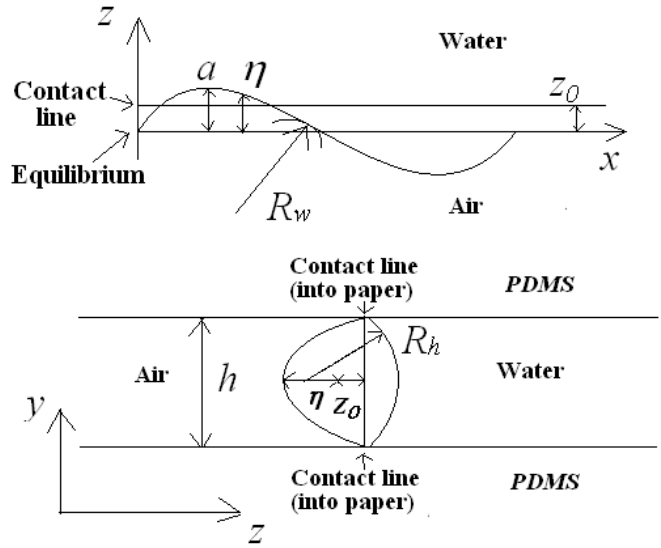

**FIG. 6**

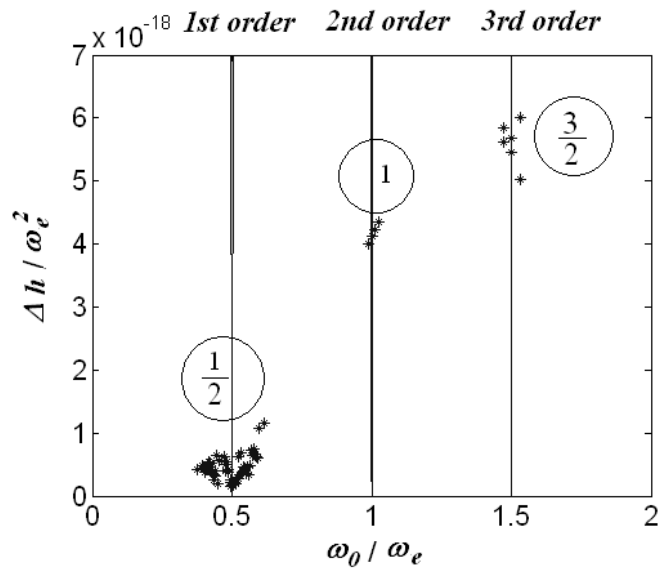

**FIG. 7**